\documentclass[aps,prd,twocolumn,superscriptaddress]{revtex4-1}

\usepackage{amsmath,setspace,amsfonts,latexsym}
\usepackage{amssymb}
\usepackage{color}
\usepackage{epsfig}
\usepackage{graphicx}
\usepackage{dcolumn}
\usepackage{bm}
\usepackage{slashed}
\usepackage{ulem}
\usepackage{array}
\usepackage[dvipsnames]{xcolor}
\usepackage[colorlinks=true,linkcolor=Blue,urlcolor=NavyBlue,citecolor=Fuchsia]{hyperref}

\definecolor{White}{rgb}{1,1,1}
\definecolor{Red}{rgb}{1,0.1,0}
\definecolor{LightYellow}{rgb}{1,1,.875}
\definecolor{SteelBlue}{rgb}{.273,.508,.703}
\definecolor{navy}{rgb}{0,0,.5}
\definecolor{LightCyan}{rgb}{.875,1,1}
\definecolor{DarkRed}{rgb}{.543,0,0}
\definecolor{HotPink}{rgb}{1,.41,.70}
\definecolor{ForestGreen}{rgb}{.13,.54,.13}
\definecolor{OliveDrab}{rgb}{.42,.55,.14}
\definecolor{MediumBlue}{rgb}{0,0,.80}
\definecolor{RoyalBlue}{rgb}{.25,.41,.88}
\definecolor{DeepSkyBlue}{rgb}{0,.746,1}
\definecolor{Brown}{rgb}{0.545,0.271,0.074}
\definecolor{Purple}{rgb}{0.637,0.285,0.641}

\def\bea{\begin{eqnarray}}
\def\eea{\end{eqnarray}}
\def\bec{\begin{center}}
\def\ec{\end{center}}

\def\beq{\begin{equation}}
\def\eeq{\end{equation}}

\newcommand\lsim{\mathrel{\rlap{\lower4pt\hbox{\hskip1pt$\sim$}}
    \raise1pt\hbox{$<$}}}
\newcommand\gsim{\mathrel{\rlap{\lower4pt\hbox{\hskip1pt$\sim$}}
    \raise1pt\hbox{$>$}}}
\def\bea{\begin{eqnarray}}
\def\eea{\end{eqnarray}}
\def\ba{\begin{array}}
\def\ea{\end{array}}
\def\bc{\begin{center}}
\def\ec{\end{center}}
\def\nn{\nonumber}

\def\fb{\rm{fb}}
\def\ab{\rm{ab}}

\newcolumntype{C}{>{\centering\arraybackslash}m{6em}}

\newcommand{\mL}{\mathcal{L}}

\begin{document}

\title{\Large Boosting Invisible Higgs Searches \\
by Tagging a Gluon Jet for Gluon Fusion Process}

\author{Won Sang Cho}
\email{wscho@snu.ac.kr}
\author{Hyung Do Kim}
\email{hdkim@phya.snu.ac.kr}
\author{Dongsub Lee}
\email{dongsub93@snu.ac.kr}
\affiliation{Center for Theoretical Physics, Department of Physics and Astronomy, \\
Seoul National University, Seoul 08826, Korea}
\begin{abstract}
We propose a novel method in that quark-gluon tagging of the jets emitted as initial state radiation (ISR) can boost searches of invisible Higgs from gluon fusion processes against irreducible electroweak vector boson productions. While quark ISR typically takes up a dominant portion than gluon in the background processes mainly by frequent quark-gluon initiated hard scatterings at the LHC, gluon ISR portion in the gluon fusion can be significantly larger in the central region of detector. Focusing on invisible Higgs searches using jet substructure variables capturing the new features, we demonstrate that Higgs from gluon fusion constrains invisible Higgs decays the most, over vector boson fusion traditionally known as the most constraining, and the limit on the branching ratio is significantly improved. 
We summarize with emphasizing that our method has wider implications in search for new resonances from gluon fusion processes.
\end{abstract}

\maketitle
%
\section{Introduction}
\label{sec:intro}

Higgs discovery at the LHC in 2012 completed the Standard Model as a description of nature in terms of elementary particles and their interactions \citep{Englert:1964et,*Higgs:1964ia,*Higgs:1964pj, Aad:2012tfa,*Chatrchyan:2012xdj,*Chatrchyan:2013lba}, 
and the precision measurement of the SM Higgs couplings is one of the most important tasks for probing new physics and the dynamics of electroweak symmetry breaking of the universe at future collider experiments \cite{Cheung:2013kla,*Cepeda:2019klc,*deBlas:2019rxi}. However Higgs precision measurements are highly non-trivial tasks in the existence of huge irreducible backgrounds.
In particular, production of electroweak vector bosons(EWVBs), such as $W$, $Z$ and $\gamma$, comprise a large portion of the irreducible background.
This is because,
1) massive gauge bosons, $Z$ and $W$, are in the mass scales similar with the Higgs, 
2) decayed particle contents are the same (or easy to be mis-identified) with the Higgs decays ($H\rightarrow f\bar{f}$), and 
3) Higgs also decays to a pair of EWVBs with $BR(H\rightarrow VV)\sim23\%$. 

Among various production mechanism of Higgs at the LHC, Higgs production via gluon fusion (ggH) \cite{[{(LO in QCD) }] Georgi:1977gs,[{(NLO in QCD) }] Ellis:1987xu,*Dawson:1990zj,*Djouadi:1991tka,*Spira:1995rr} has the most dominant contribution ($90\%$) to the total production cross sections. 
The ggH process is a very unique process in that it can transform between the state of QCD force carriers and electroweak bosons via quark loops, and not leaving any other QCD remnants at leading order (LO), so its event topology can basically be the same with the EWVB productions from the leading orders.
In result, tagging the Higgs from gluon fusion has been suffering from the irreducible backgrounds much more than the other sub-dominant productions including vector boson fusion (VBF), Higgsstrahlung (VH), and $t\bar{t}H$, as it does not have associated objects with fixed particle identity good for tagging the whole process.

For this reason the most stringent constraints for probing the Yukawa couplings of Higgs have usually been obtained via the non-ggH processes, e.g. in $H\rightarrow$ $b\bar{b}$ \cite{Aaboud:2018zhk,*Sirunyan:2018koj}, $c\bar{c}$ \cite{Aaboud:2018fhh}, $\tau^+\tau^-$ \cite{Sirunyan:2018cpi,*Aaboud:2018pen,*Sirunyan:2017khh}, $\mu^+\mu^-$ \cite{Sirunyan:2018hbu,*Aaboud:2017ojs}, $e^+ e^-$ \cite{Khachatryan:2014aep,*Aad:2019ojw}.
The same argument also applies to the searches of Higgs pair production via gluon fusion against the EWVB backgrounds, but in this case things can get worse as the dominant ggH contribution increases ($93\%$).

In this paper we revisit and generalize an overlooked property, and investigate a new possibility for boosting Higgs searches via ggH.
We focus on the sizeable differences in quark-gluon composition of the central ISR jets between the general ggH productions and their irreducible EWVB backgrounds.
Based on such difference, we then show that {\it the tagging the central gluon jets from ISR can provide useful discrimination power to overall Higgs searches.}
The difference was stated earlier in \cite{Choudhury:1993hv} without attention, 
and the new possiblity on the difference was claimed in \cite{Lee:2018jxx} for $H\rightarrow \mu\mu$, 
and studied \cite{Kasieczka:2018lwf} for a monojet analysis. 
As the new method can have big impacts,
here we generalize the property for (multiple) Higgs and EWVB productions, 
emphasizing that in the central region of detector, leading ISR jet from ggH is mostly a gluon jet.
To prove its experimental feasibility, then {\it we apply the new method in search for invisible Higgs decays, and show that the limit on the Higgs invisible decay branching ratio can be improved significantly ($60\%\rightarrow 5\%$) for the most dominant gluon fusion, to be the most constraining channel,
which has been not so useful compared to the other channels at the LHC.}

This paper is composed as the following.
	In section~\ref{sec:kin_leading_jet}, we discuss the dynamics of the ISR from two main processes, gluon fusion Higgs production and massive vector boson production, quantitatively at the leading order, and investigate on the existence of gluon-enriched kinematic region of leading ISR jet from the ggH process.
The discussion is generalized by including the case of general Higgs signal production with $n$ Higgs ($\rm ggH^{n}$+jets) and EWVB production ($V^{n}$+jets).
In section~\ref{sec:invHiggs} we demonstrate the performance of the event classifier based on the ISR jet properties at detector level, in search for invisible Higgs decay.
The performances of various discrimination models are compared with in terms of the upper limit of confidence level on invisible Higgs branching ratio in section~\ref{sec:mva_result}, and section~\ref{sec:conclusion} is devoted to conclusion.
Miscellaneous details about multivariate models using deep neural networks and the others are added in Appendix.

%
\section{Dynamics of Leading ISR Jet}\label{sec:kin_leading_jet}

\begin{figure}[!tbp]
	\includegraphics[width=0.45\textwidth]{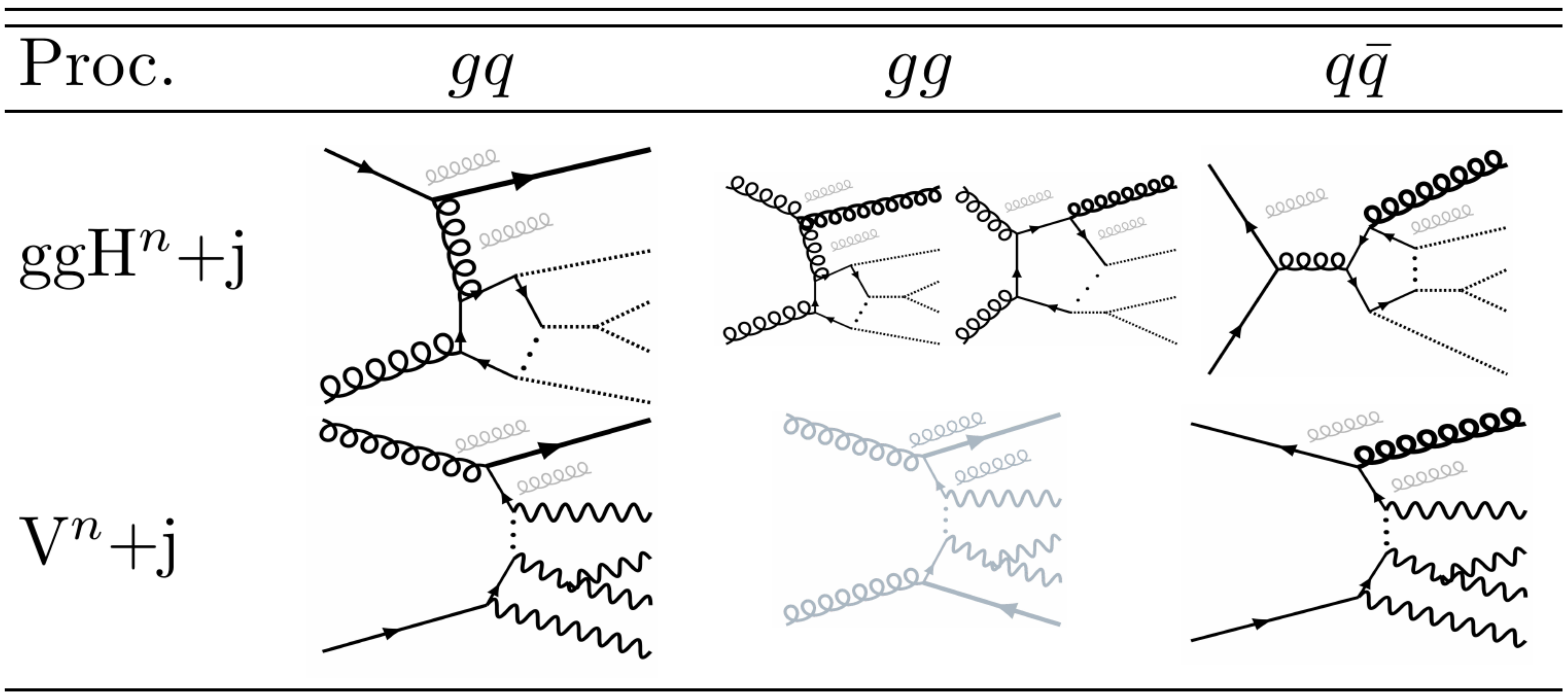}
	\caption{Leading diagrams (bold) of the (multi) Higgs productions from gluon fusion (ggH$^n$+jets), against the corresponding irreducible (multi) EWVB backgrounds (V$^n$+jets) with additional ISR(s) for 3 parton initial states($gq$, $gg$ and $q\bar{q}$). } 
	\label{fig:diagrams}
\end{figure}

In this section, we discuss about the dynamics of ISR at leading order associated with the ggH and EWVB productions at the LHC.
Fig.~\ref{fig:diagrams} shows the leading diagrams (in bold) of general Higgs signal productions from gluon fusion (ggH$^{n}$+jets) and EWVB productions (V$^n$+jets) as irreducible backgrounds with an emission of ISR(s), for three different initial parton configurations ($gq$, $gg$, $q\bar{q}$). 
Here the $n$ can be larger than one for multi Higgs/EWVB productions, and additional Higgs or EWVB productions with $n\ge 2$ is also represented.
The gluon lines in grey indicate extra gluon emissions and the V$^n$+jets diagram (bottom-center) from $gg$ initial states is drawn also in grey as it is subleading to the other 5 diagrams in $\alpha_s$.
By the irreducibility we can assume that the particle IDs from the decays of the Higgs and EWVB are the same or very similar. 

\begin{figure*}[!tbp]
	\includegraphics[width=0.27\textwidth]{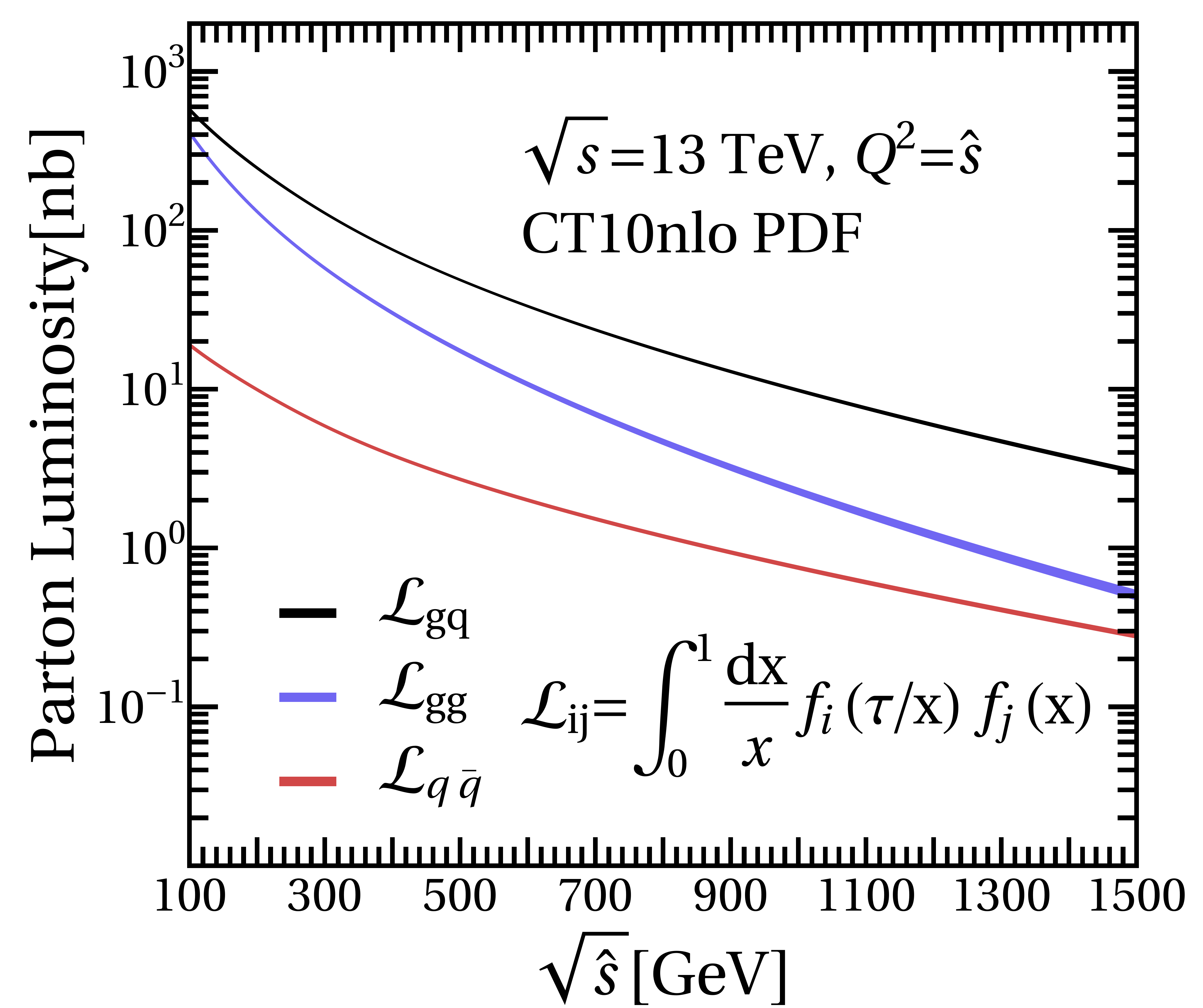} \hspace{1cm}
	\includegraphics[width=0.27\textwidth]{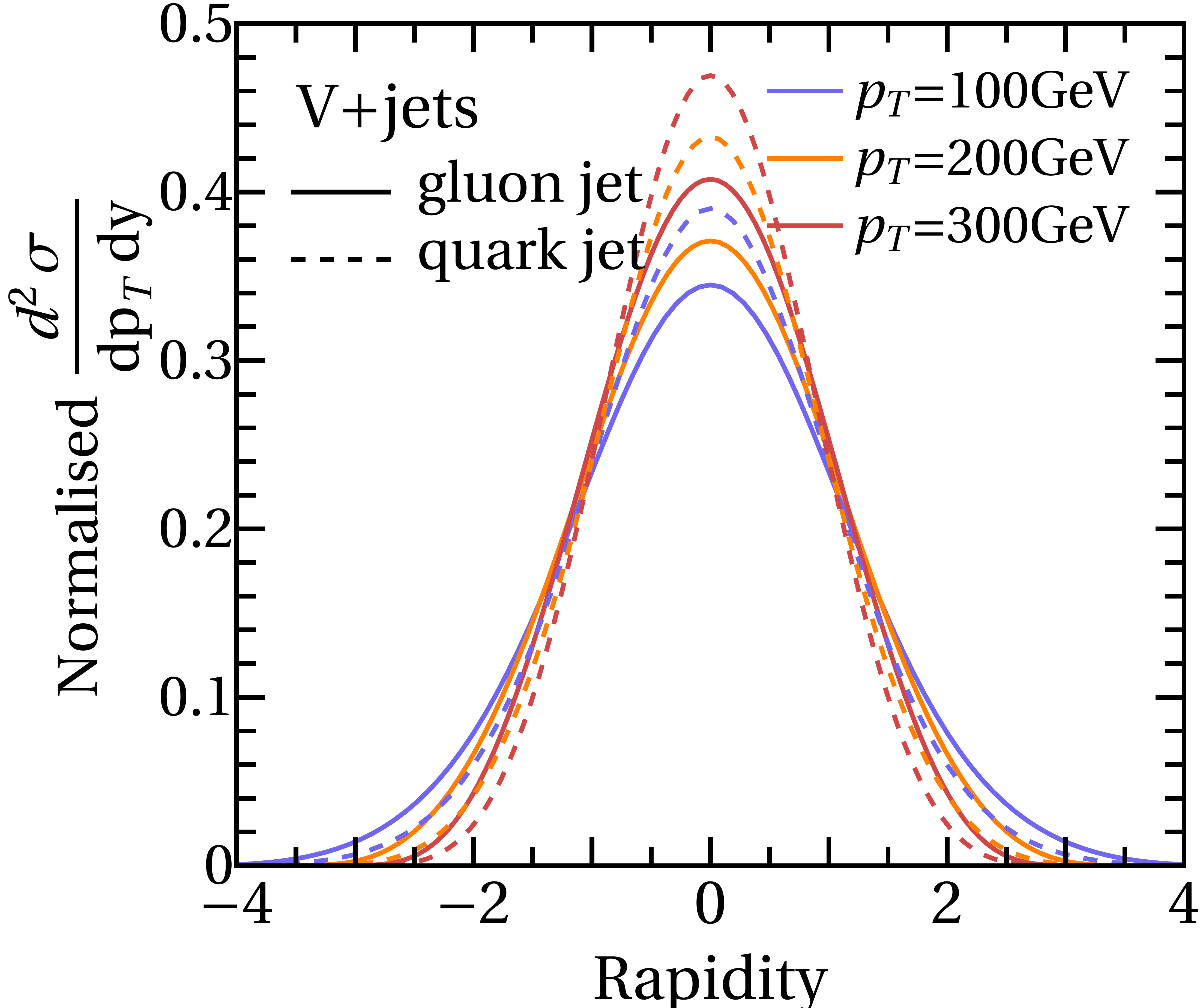}\hspace{1cm}
	\includegraphics[width=0.27\textwidth]{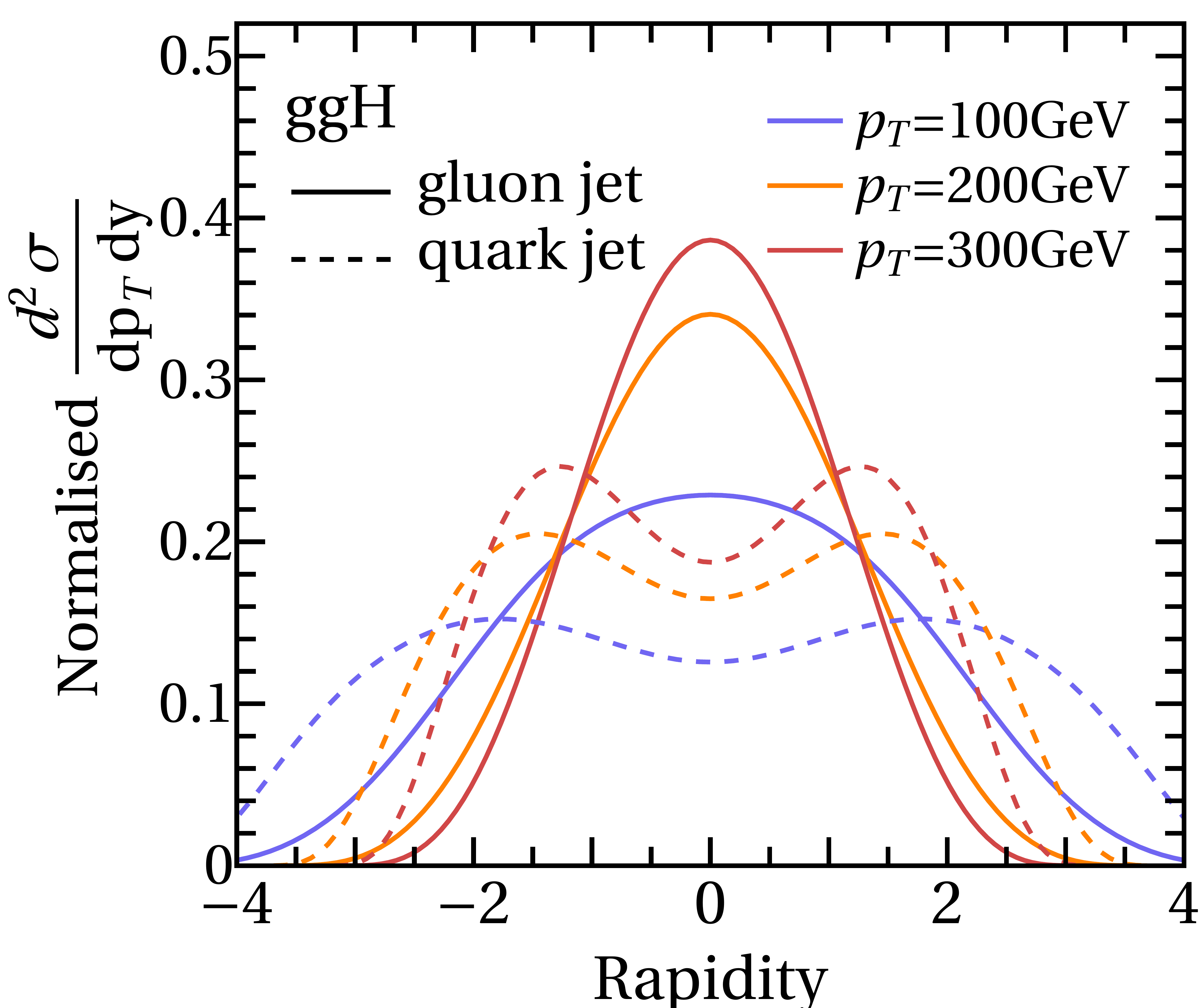}
	\caption{(a)The parton luminosity $\mathcal{L}_{ij}$ with parton density function(PDF) $f_{i}(x,Q^2=\hat{s})$ at the LHC with CT10nlo PDF as indicated in the figure. (b)Rapidity distribution of leading gluon and quark ISR jets from $Z$ boson production and (c)from ggH processes.}
	\label{fig:lhc_lumi_rapidity}
\end{figure*} 

From the Fig.~\ref{fig:diagrams}, it should be first noted that the flavor of ISR(s) emitted in the leading diagrams (in bold) are uniquely fixed since the other final state (H$^{n}$/V$^{n}$) accompanied with is a colorless non-QCD particle in the 2 to (n+1) processes, for the given initial parton configurations. 
It is also noticeable that the parton luminosity functions, $\mathcal{L}_{gq,gg,q\bar{q}}$ referring to Fig.~\ref{fig:lhc_lumi_rapidity}(a), for the three initial partonic states are hierarchical - $\mathcal{L}_{gq}$ $>$ $\mathcal{L}_{gg}$ $\gg$ $\mathcal{L}_{q\bar{q}}$, e.g. $\mathcal{L}_{gq}$:$\mathcal{L}_{gg}$:$\mathcal{L}_{q\bar{q}}$ $\sim$ $2$:$1$:$0.07$ at $\sqrt{\hat{s}}\sim 100$ GeV, and such a hierarchy persists to higher energy scale as shown in the same plot.
Based on these two observations, the dominant flavor of leading ISR jet from the whole ggH$^{n}$+jets and V$^n$+jets processes can be predicted and their quark-gluon compositions can be compared.

\begin{figure}[!tbp]
\centering
\vspace*{-0.3cm}
	\includegraphics[width=0.35\textwidth]{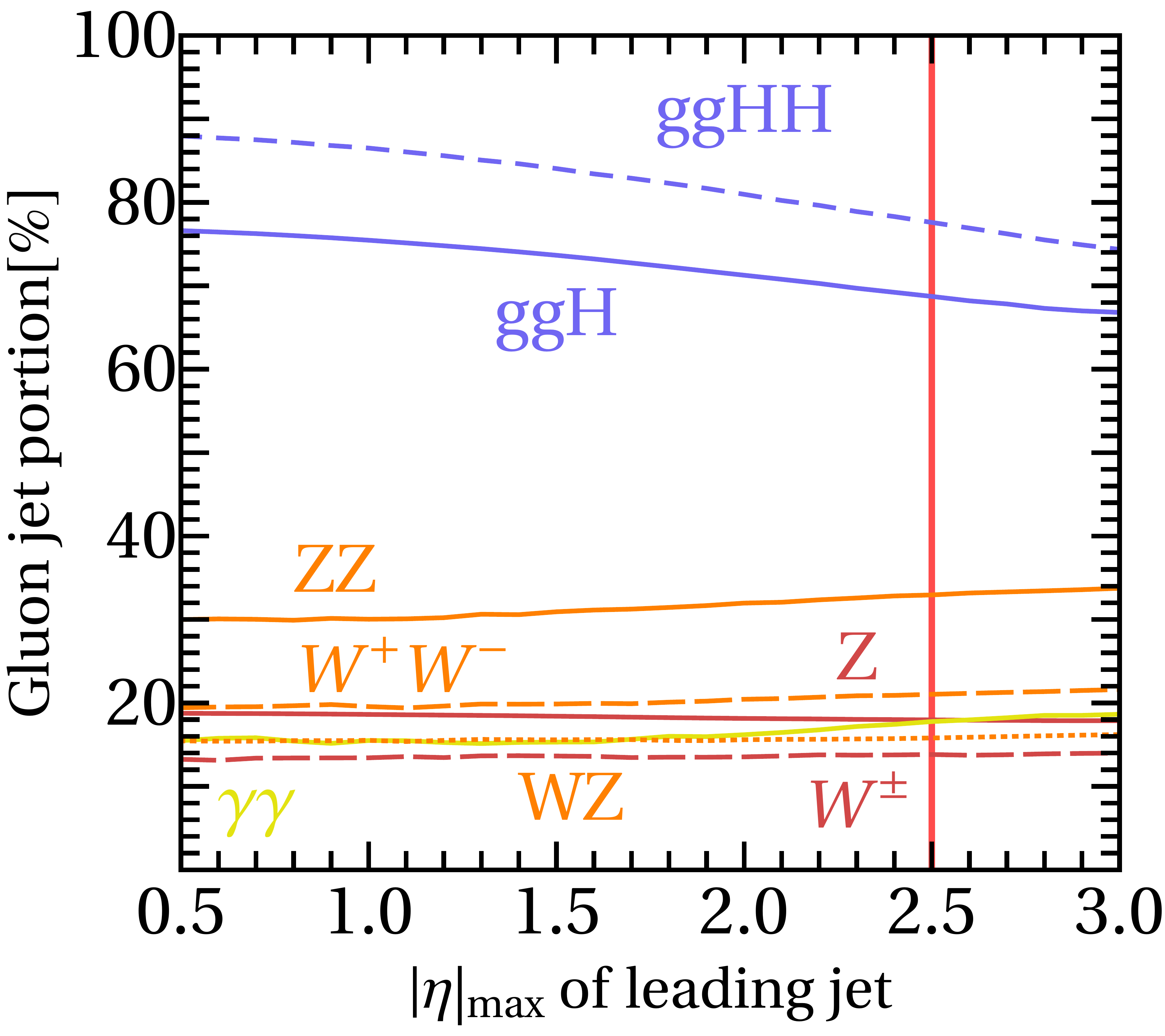} 
	\caption{Gluon portion over $|\eta^{j_1}|_{max}$ ($p^{j_{1}}_{T}$ $>100$ GeV) of the leading ISR jet associated with ggH$^n$ and V$^n$ productions.}
	\label{fig:parton_contents}
\end{figure}

Let us consider the background process, the production of EWVB, first.
In $n=1$ case, the differential partonic cross sections in the center-of-mass frame with respect to Mandelstam variable $\hat{t}$ are given as the following for the respective processes, $q\bar{q} \to Vg$ and $gq \to Vq$:
\begin{eqnarray}
	\frac{d\hat{\sigma}_{Vg}}{d\hat{t}}  &=& \frac{\alpha_{3}}{16 \hat{s}^{2}}\frac{C_{F}}{N_c}\sum_{q}{\left( {g^{q}_{V}}^{2}+{g^{q}_{A}}^{2} \right)\frac{\hat{t}^{2}+\hat{u}^{2}+2\hat{s}m^{2}_{V}}{\hat{t}\hat{u}}},  \\
	\frac{d\hat{\sigma}_{Vq}}{d\hat{t}} &=&  \frac{\alpha_{3}}{16 \hat{s}^{2}}\frac{T_{F}}{N_c}\sum_{q}{\left( {g^{q}_{V}}^{2}+{g^{q}_{A}}^{2} \right)\left( -\frac{\hat{s}^{2}+\hat{u}^{2}+2\hat{t}m^{2}_{V}}{\hat{s}\hat{u}}\right)},\nonumber
\end{eqnarray}
with the number of colors $N_c$, the mass of gauge boson $m_{V}$, strong coupling $\alpha_{3}$, vectorial coupling $g^{q}_{V}$, and axial coupling $g^{q}_{A}$ between $V$ and massless quarks $q$, so e.g. for $Z$-boson case, $g^{q}_{V}$ $=$ $\frac{g_{2}}{\cos\theta_{W}}\left(\frac{1}{2}T^{q}_{3}-Q^{q}\sin^{2}\theta_{W} \right)$ and $g^{q}_{A} = \frac{g_{2}}{2\cos\theta_{W}}T^{q}_{3}$.
The $T_{F}$(=1/2) and $C_{F}$(=4/3) are Dynkin index and quadratic Casimir of fundamental representation of QCD.
Since the matrix elements corresponding to the cross sections have crossing symmetry with each other, 
so their angular profiles are not so distinctive in the LAB frame even after the integration with parton distribution functions
as can be seen in Fig.~\ref{fig:lhc_lumi_rapidity}(b).\\ 
\indent However their total cross sections become very different considered with the two main scaling factors 
- 1) different interaction strengths from different initial states, and 2) hierarchical initial parton luminosity function values.
As the $q\bar{q} \to V g$ process is averaged over two quarks, so the amplitude has $\text{tr}(t_{a}t^{a})/N^{2}_{c}$=$C_{F}/N_{c}$ factor with $t_{a}$ as the $SU(3)_{c}$ generators at fundamental representation,
while the $g q \to V q$ process has $\text{tr}(t_{a}t^{a})/(N_{c}(N^{2}_{c}-1)) = T_{F}/N_{c}$.
Although $T_{F} < C_{F}$, the hierarchically larger parton luminosity function of $gq$, $\mathcal{L}(gq)$ compared to $\mathcal{L}(q\bar{q})$, makes the hadronic total cross section $\sigma(p p \to V q)$ much bigger than the $\sigma(p p \to V g)$ by $\mathcal{O}(10)$ at the LHC.\\
\indent This property can be generalized also for V$^n(\ge2)$+jets processes, 
and in result the leading ISR jet is expected most likely to be a quark jet for general EWVB processes.
As a demonstration of the quark jet portion dominance, we checked gluon jet portions, $R^g$, of the leading ISR ($|\eta^{j_1}|\le 1$ and $p^{j_1}_T$ $>100$ GeV) in various EWVB processes, and they are found to be 
(1) $R^{g}_{(WW,\,WZ,\,ZZ)}$ $\approx$ ($0.20$,$0.16$,$0.30$) for V$^{n=2}$+jets processes, 
(2) $R^{g}_{(W,Z)}$ $\approx$ ($0.13$,$0.19)$) for V$^{n=1}$+jets, and (3) $R^{g}_{\gamma\gamma}$ $\approx$ $0.15$ for prompt di-photon+jet processes, as in Fig.~\ref{fig:parton_contents}.

For the ggH signal, since $\mL_{q\bar{q}}$ is much smaller than the others, $q\bar{q} \to H g$ process can be ignored for the estimation.
Relevant partonic differential cross sections, $d\hat{\sigma}_{Hq}/d\hat{t}$ from $g q \to H q$ and $\hat{\sigma}_{Hg}/d\hat{t}$ from $gg \to Hg$ process, computed at NLO with massless five quarks and finite top-mass effect, 
	are given \cite{Ellis:1987xu,*Dawson:1990zj,*Djouadi:1991tka,*Spira:1995rr} as the following:
\begin{eqnarray}
	\frac{d\hat{\sigma}_{Hg}}{d\hat{t}} &=& \frac{ \alpha^{3}_{3}}{16\pi^{2}\hat{s}^{2} v^{2}}\frac{C_{A}}{N^{2}_{c}-1}\frac{m^{8}_{H}}{\hat{s}\hat{t}\hat{u}} \left( \left| A_{2}(\hat{s},\hat{t},\hat{u})\right|^{2}\right. \\
				&\,& \left. + \left| A_{2}(\hat{t},\hat{u},\hat{s})\right|^{2} + \left| A_{2}(\hat{u},\hat{s},\hat{t})\right|^{2} + \left| A_{4}(\hat{s},\hat{t},\hat{u})\right|^{2}\right), \nn \\
	\frac{d\hat{\sigma}_{Hq}}{d\hat{t}} &=& \frac{ \alpha^{3}_{3}}{64\pi^{2}\hat{s}^{2} v^{2}}\frac{C_{F}}{N^{2}_{c}-1}\left( -\frac{\hat{s}^{2}+\hat{u}^{2}}{\hat{t}} \right) \frac{m^{4}_{H}|A_{5}(\hat{t},\hat{s},\hat{u})|^{2}}{(\hat{s}+\hat{u})^{2}}. \nn
\end{eqnarray}
Here $m_{H}$ is mass of Higgs, and $C_{A}$ is quadratic Casimir of adjoint representation of $SU(3)_{c}$ so that $C_{A} = N^{2}_{c}-1$ is 8.
$A_{2}$, $A_{4}$ and $A_{5}$ are loop functions of which definitions are in Appendix~\ref{app:higgs_loop-functions}.
From $\mL_{gg}< \mL_{gq}$, one may think that the hadronic cross section $\sigma(pp \to Hq)$ is larger than the other one, $\sigma(pp \to Hg)$, but there is a difference in associated color factors.
It makes $C_{A}/C_{F} = 9/4$ enhancement on $\sigma(pp \to Hg)$, and two cross sections are compatible with each other.
Therefore the quark-gluon portion of leading ISR jet can highly depend on the dynamics of the leading ISR from the two dominant signal processes, especialy on their transverse momentum and angular distributions.\\
\indent To understand the differences on the $p_T$ profiles of the two processes, 
let's consider the infinite-top-mass ($m_{t}\to\infty$) limit.
This limit is safe as long as $|\hat{s} - m^{2}_{H}|/(4m^{2}_{t}) \lesssim 1 $, with mass of top-quark, $m_{t}$.
In this limit, loop functions are approximated as
\begin{equation}
	A_{2}(\hat{s},\hat{t},\hat{u}) \to - \frac{\hat{s}^{2}}{3m^{4}_{H}}, \quad A_{4} \to - \frac{1}{3}, \nn
\end{equation}
and
\begin{equation}
	A_{5}(\hat{s},\hat{t},\hat{u}) \to \frac{2}{3} \frac{\hat{t}+\hat{u}}{m^{2}_{H}}, \nn
\end{equation}
also for all the other permuted arguments.
The effective differential cross sections, $d\hat{\sigma}^{\rm HEFT}_{Hg}/d\hat{t}$ and $d\hat{\sigma}^{\rm HEFT}_{Hq}/d\hat{t}$ are
\begin{eqnarray}
	\frac{d\hat{\sigma}^{\rm HEFT}_{Hg}}{d\hat{t}} &=& \frac{ \alpha^{3}_{3}}{144\pi^{2}\hat{s}^{2} v^{2}}\frac{C_{A}}{N^{2}_{c}-1} \frac{\hat{s}^{4}+\hat{t}^{4}+\hat{u}^{4}+m^{8}_{H}}{\hat{s}\hat{t}\hat{u}} \\
	\frac{d\hat{\sigma}^{\rm HEFT}_{Hq}}{d\hat{t}} &=& \frac{\alpha^{3}_{3}}{144\pi^{2}\hat{s}^{2} v^{2}}\frac{C_{F}}{N^{2}_{c}-1} \left( - \frac{\hat{s}^{2}+\hat{u}^{2}}{\hat{t}}\right). \nn 
\end{eqnarray}
Note that, each Mandelstam variable at the lowest order of transverse momentum $p_{T}$ is given by, $\hat{s} \simeq m^{2}_{H}$,
\begin{equation}
	\hat{t} \simeq m_{H}p_{T}e^{\Delta y} \quad \text{and} \quad \hat{u} \simeq m_{H}p_{T}e^{-\Delta y}, \nn
\end{equation}
where $\Delta y \equiv y_{b} - y_{j}$ is the rapidity difference of the two initial parton system ($y_{b}$) and the jet ($y_{j}$) in the LAB frame.
Therefore,
\begin{equation}
	\frac{d\hat{\sigma}^{\rm HEFT}_{Hg}}{d\hat{t}} \propto \frac{1}{p^{2}_{T}} \quad \text{and} \quad \frac{d\hat{\sigma}^{\rm HEFT}_{Hq}}{d\hat{t}} \propto \frac{1}{p_{T}},
	\label{eq:pt}
\end{equation}
which implies that a gluonic leading ISR jet is likely to have softer $p_T$ than a quark ISR jet.\\
\indent The angular dependence can also be figured out by considering $s$-wave scattering of two processes.
If the orbital angular momentum is zero, the spin of outgoing gluon should be aligned along the direction perpendicular to the beam, otherwise the total angular momentum is not conserved.
However, the outgoing quark is likely to be backward to the incoming quark for angular momentum conservation in $g q \to Hq$.
Then, combined with the balanced momentum profile of the initial $gg$ state at the LHC, the gluon jets are likely to be emitted more in central rapidity region with soft $p_T$, in comparison to the quark jet as can be seen in Fig.~\ref{fig:lhc_lumi_rapidity}(c).\\
\indent This property can also be applied to ggH$^n(\ge2)$+jets processes,
and in result the leading ISR jet is expected likely to be a gluon jet for general ggH processes.
As a demonstration of the gluon jet portion dominance, 
we also showed the gluon jet portions, $R^g$, of the leading ISR ($|\eta^{j_1}|\le 1$ and $p^{j_1}_T$ $>100$ GeV), 
associated with a single (ggH) and a pair of Higgs (ggHH) productions from gluon fusion. 
They are found to be 
(1) $R^{g}_{ggH}$ $\approx$ $0.75$ for ggH$^{n=1}$+jets, and 
(2) $R^{g}_{ggHH}$ $\approx$ $0.87$ for a pair of Higgs production, 
as in Fig.~\ref{fig:parton_contents}.
If $P_T$ cut is lowered to $P^{j_1}_{T} > 50$ GeV, the $R^{g}$ for the ggH$^{n}$(V$^{n}$) process increases(decreases), respectively, by $\sim 2$-$5\%$ in the $|\eta^{j_1}|_{max}$ range of Fig.~\ref{fig:parton_contents}, 
which is also consistent with our expectation of the soft $p_T$ dominance of gluon ISR from the ggH processes, as in Eq.~\ref{eq:pt}.   

\begin{figure*}[!t]
\centering
		\includegraphics[width=0.24\textwidth]{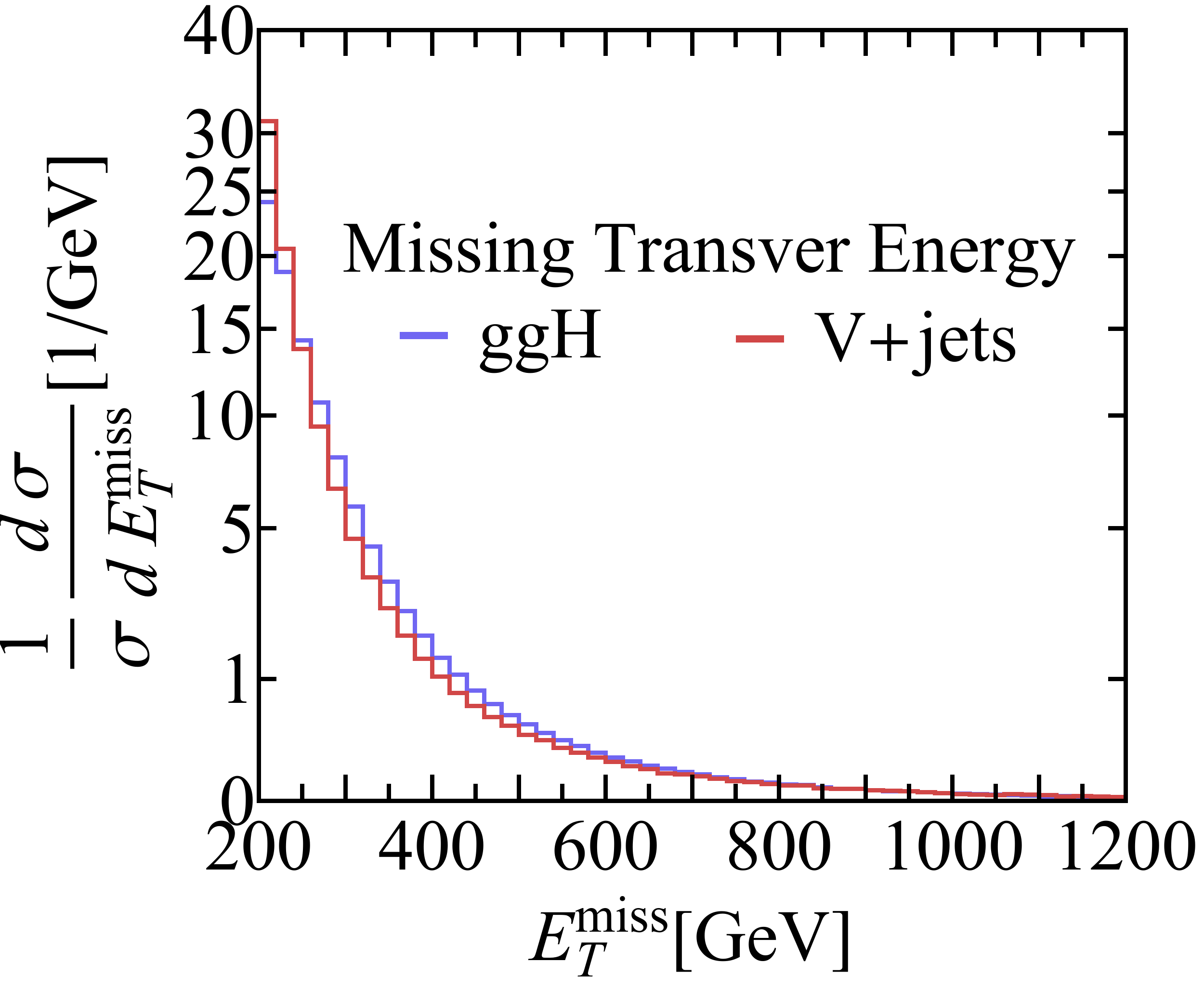}
		\includegraphics[width=0.24\textwidth]{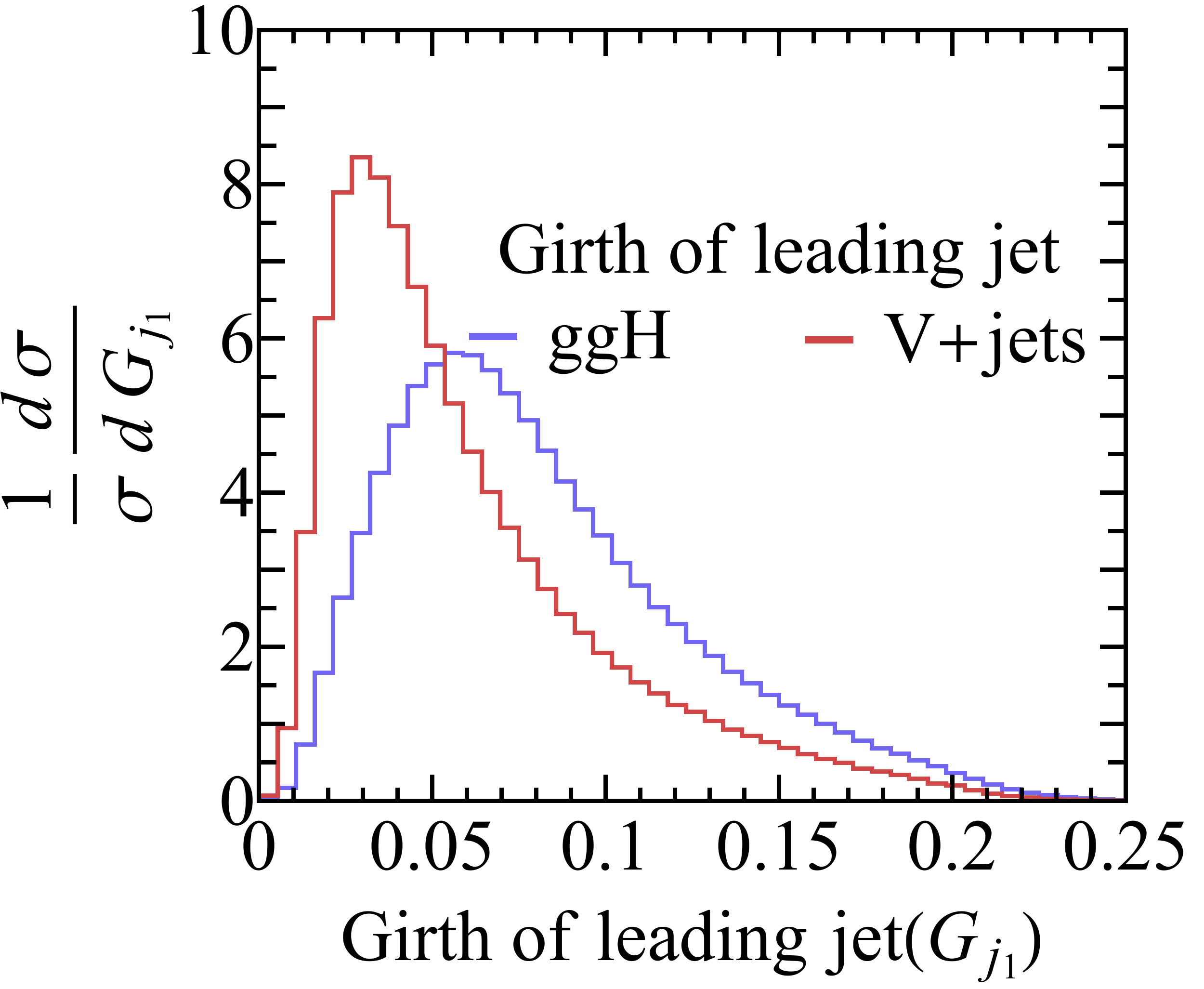}
		\includegraphics[width=0.24\textwidth]{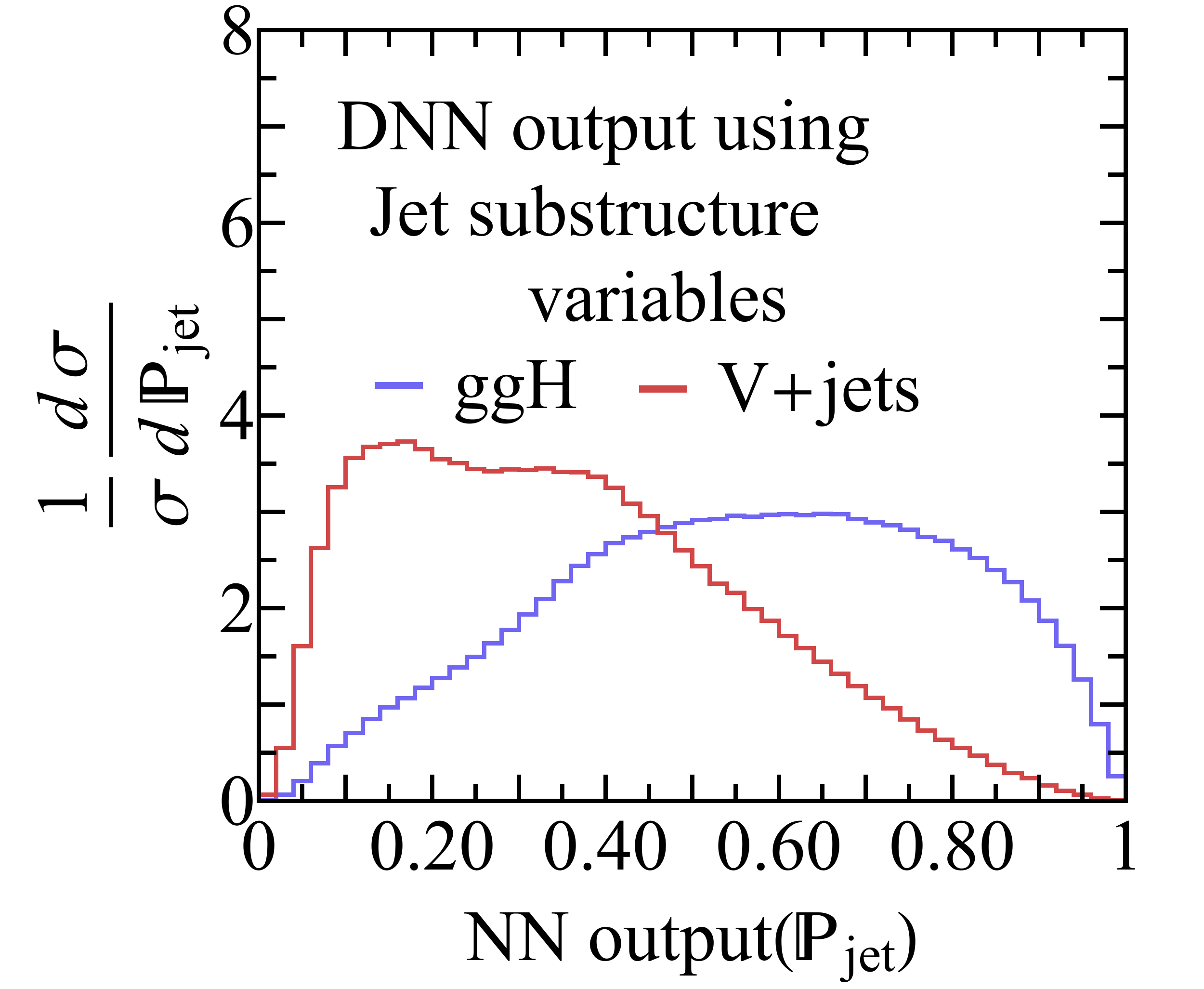}
		\includegraphics[width=0.24\textwidth]{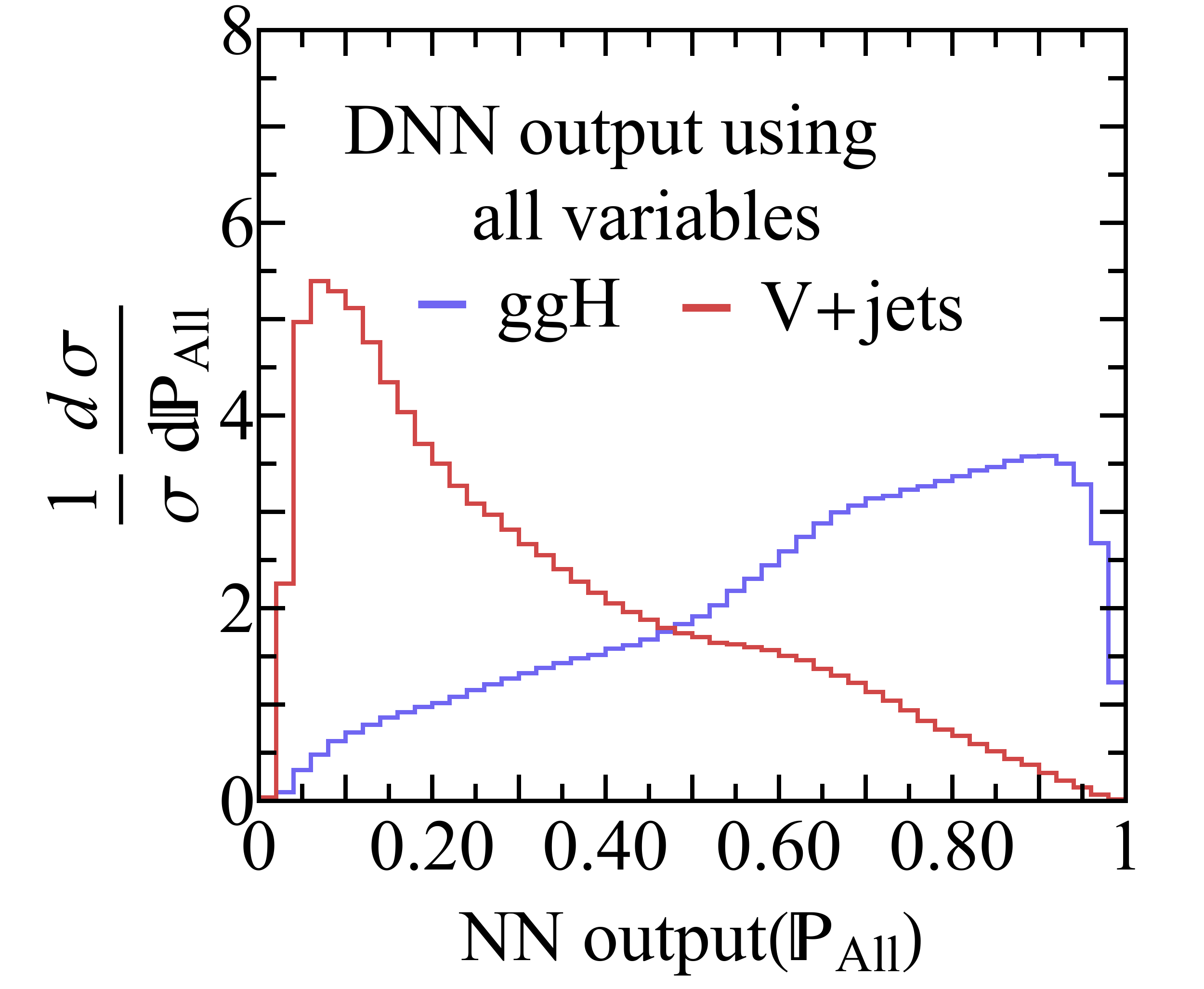}
	\caption{Signal and background profiles in various templates, (a) $E^{\text{miss}}_{T}$, (b) Girth of leading jet, (c) event classifier $\text{P}_{\text{S/B}}(S^{\,\text{jet}})$ (1:ggH-like, 0:V+jet) trained using the jet substructure observables $S^{\,\text{jet}}$, and (d) $\text{P}_{\text{S/B}}$ $(S^{\,\text{jet}} \cup$  $\{\text{kin.}\})$ using all features.}
	\label{fig:templates}
\end{figure*}

Like as the two quark jets in forward region from the VBF, and as the extra Z/W from the VH, now the ggH$^n$ + jets also has such a unique property accompanied - the {\it gluon-like ISR jets in central region}.
In this regard, if some relevant techniques using quark-gluon tagging of ISR jets are employed, one can improve the constraints from the most dominant gluon fusion channel for a broad range of Higgs signatures which are buried in the irreducible EWVB backgrounds matched with.

%
\section{Invisible Decay of Higgs}\label{sec:invHiggs}

In order to demonstrate experimental feasibility, here we utilize the gluonic ISR jet from the ggH for constraining {\it invisible decays of Higgs} where the decayed particles from the Higgs and irreducible EWVB are electrically neutral and invisible.
Historically, there had been lots of studies on the possibility of invisible Higgs decays along the developements of the Standard Model and the beyonds.
The early proposals include the models in diversity e.g. with
Majorons\cite{Chikashige:1980ui,*Gelmini:1980re,Shrock:1982kd,*Li:1985hy,*Jungman:1991sh,*Joshipura:1992ua,*Joshipura:1992hp},
supersymmetries \cite{Griest:1987qv,*Gunion:1988yc,*Romao:1992zx,*deCampos:1995ten,*Djouadi:1996mj,*Djouadi:1997gw,*Belanger:2001am,*Hirsch:2004rw,*Antoniadis:2004se,*Chang:2005ht,*Draper:2010ew,*Cao:2012im,*Butter:2015fqa,*Barman:2017swy,*Wang:2020dtb,*Barman:2020vzm}, 
heavy neutrinos with radiatively generated masses \cite{Pilaftsis:1991ug}, 
large extra dimensions \cite{ArkaniHamed:1998rs,Dienes:1998sb,*ArkaniHamed:1998vp,*Giudice:2000av,*Battaglia:2004js,*Datta:2004jg,*Dominici:2009pq}, 
the 4th generations \cite{Fargion:1999ss,*Khoze:2001ug,*Belotsky:2002ym,Belotsky:2004ex,Frere:2004rh,*Kribs:2007nz,*Eberhardt:2012sb}, and so on, 
while the recent interpretations are mainly based on the effective singlet extensions of the SM, 
in the context of so-called Higgs-portal models with dark matters\cite{Silveira:1985rk,*Krasnikov:1992zk,*McDonald:1993ex,*Binoth:1996au,*Datta:1997fx,*Burgess:2000yq,*Bento:2001yk} 
\cite{Davoudiasl:2004be,*Schabinger:2005ei,*Barbieri:2005ri,*Patt:2006fw,*Cerdeno:2006ha,
*OConnell:2006rsp,*Kim:2006af,*BahatTreidel:2006kx,
*Barger:2007im,*Bertolami:2007wb,
*Kim:2008pp,*Barger:2008jx,
*Gonderinger:2009jp,*Cheung:2009wb,*He:2009yd,
*Andreas:2010dz,*Kanemura:2010sh,
*Englert:2011yb,*Raidal:2011xk,*He:2011de,*Pospelov:2011yp,*Lebedev:2011iq,*Baek:2011aa,*Drozd:2011aa,*Djouadi:2011aa,*Batell:2011pz,*He:2011gc,*Kamenik:2012hn} 
\cite{Djouadi:2012zc,Cline:2013gha,*deSimone:2014pda,*Baek:2014jga,*Feng:2014vea,*Beniwal:2015sdl,*Han:2016gyy,*Dupuis:2016fda,*Arcadi:2019lka}. 
For these proposals, there exist numerous phenomenological and experimental researches in search for the invisible Higgs decays,
via the production channels, including VH \cite{Choudhury:1993hv,Frederiksen:1994me,Davoudiasl:2004aj,*Martin:1999qf,*Godbole:2003it,Englert:2011us,Bai:2011wz,Ghosh:2012ep,Carpenter:2012rg,*Goncalves:2016bkl,*Goncalves:2018ptp} 
\cite{Aad:2013oja,*Aad:2014iia,*Aad:2014vka,*Aad:2015uga,*CMS:2016pps,*Aaboud:2017bja,*Aaboud:2018xdl,Khachatryan:2016whc,Sirunyan:2017jix,Chatrchyan:2014tja,ATLAS:2018tor,Aaboud:2019rtt},
$t\bar{t}H$ \cite{Gunion:1993jf} \cite{Choudhury:1993hv,Kersevan:2002zj,*Malawski:2004kg,*Zhou:2014dba,*Craig:2015jba,*Buckley:2015ctj} 
\cite{CMS:2019bke},
ggH (monojet)
\cite{Choudhury:1993hv,Davoudiasl:2004aj,Fox:2011pm,Englert:2011us,Bai:2011wz,Djouadi:2012zc,Barducci:2016fue} 
\cite{Chatrchyan:2011nd,*Aad:2011xw,*Aaltonen:2012jb,*Chatrchyan:2012me,*ATLAS:2012zim,*ATLAS:2012ky,*Khachatryan:2014rra,
*Aad:2015zva,*Aaboud:2016tnv,*Sirunyan:2017hci,Khachatryan:2016whc,Sirunyan:2017jix},
and 
VBF \cite{Eboli:2000ze} \cite{Davoudiasl:2004aj,Bai:2011wz,Ghosh:2012ep,Bernaciak:2014pna,*Goncalves:2017gzy,*Biekotter:2017gyu,*Heisig:2019vcj} \cite{DiGirolamo:2002vwa,*Aad:2015txa,*Aad:2015pla,*CMS:2016jjx,*Aaboud:2018sfi,*ATLAS:2020cjb,Khachatryan:2016whc,Chatrchyan:2014tja,ATLAS:2018tor,Sirunyan:2018owy,Aaboud:2019rtt} which has been presented the most sensitive limits on the invisible Higgs decay BR at the LHC.
There also have been interesting surveys via diffractive Higgs productions \cite{Belotsky:2004ex}, total decay width \cite{Low:2011kp}, 
Higgs rare B decays \cite{Kim:2009qc}, di-Higgs \cite{Banerjee:2016nzb,*Englert:2019eyl}, Higgs off-shell decays \cite{Endo:2014cca,*Craig:2014lda,*Ruhdorfer:2019utl}, lepton colliders \cite{Eboli:1994bm,*Searches:2001ab,*Abdallah:2003ry,*Djouadi:2007ik,*Kanemura:2011nm,*Baer:2013cma,*Moortgat-Picka:2015yla,*Kato:2020pyl,*Tang:2015uha}, including global analysis \cite{Englert:2011aa,*Espinosa:2012vu,*Belanger:2013kya,*Bechtle:2013wla,*Curtin:2013fra,*Corbett:2015ksa,*Athron:2018hpc,*Arina:2019tib}.
Throughout the searches, the dominant $ggH$ channel has never been competitive to the other sub-dominant channels. However, employing the new method we show that the most stringent constraint can be obtained from the ggH channel for the invisible Higgs decays, as is demonstrated in the next paragraphs.

Assuming the Higgs production cross section of the SM, we perform the analysis in search for the invisible Higgs decays in $E_T^{\text{miss}}$+jets signature via the ggH+jets channel. Samples are generated by Monte Carlo simulated pp collisions at a center-of-mass energy of $\sqrt{s}=13$ TeV at the LHC, for 36 fb$^{-1}$ using  MadGraph5 aMC@NLO v2.6.2 \cite{Alwall:2014hca} interfaced with Pythia v8.235 \cite{Sjostrand:2014zea} for hadronisation and fragmentation. Delphes v.3.4.1 is used for detector simulation \cite{deFavereau:2013fsa}. 
The signal process (ggH+jets) is generated with up to extra 1 jet at LO taking into account finite top mass effects \cite{Hirschi:2015iia} with $M_H=125$ GeV, and backgrounds are generated at NLO in QCD.
We use FxFx scheme with $k_{T}$-algorithm and $\Delta R = 1$ for jet merging \cite{Frederix:2012ps}.
For jet clustering, FastJet v3.2.1 \cite{Cacciari:2011ma} is used with anti-$k_{T}$ algorithm with $\Delta R=0.4$, and CT10NLO \cite{Lai:2010vv} is used for parton distribution function.

Among the relevant background processes - $V(Z(\nu\nu)$, $W(\ell\bar{\nu}))$+jets, Diboson, top quarks, $Z/\gamma\rightarrow\ell\bar{\ell}$, QCD multijets, where the leptons ($\ell$) in $W/Z/\gamma$ decays are mis-identified, we only included the most dominant irreducible EWVB backgrounds - $V$+jets, while the others take just $O(1)\%$ level for the event selection criteria as follows \cite{[{(e.g. see Table.7 and Fig.5) }] Khachatryan:2016whc}: 
\begin{itemize}
	\item $p^{j_{1}}_{T} > 100\text{GeV}$, $|\eta^{j_{1}}| < 2.5$, $E^{\text{miss}}_{T} > 200\text{GeV}$, $\min_{j \in \{\text{jets}\}}\Delta \phi(\vec{p}^{\; \text{miss}}_{T},\vec{p}^{\; j}_{T}) \ge 0.5$, $N_{\text{jet}}\ge 1$.
\end{itemize}
The 1st (2nd) cut on the transverse momentum (pseudorapidity) of the leading jet is imposed to suppress all of the backgrounds, the 3rd cut on the missing transverse energy is mainly to reduce the QCD and top quarks, and the 4th cut with the missing transverse momentum, $\vec{p}_T^{\;\text{miss}}$ suppresses the QCD multijets very efficiently \cite{Sirunyan:2017jix}.
 
There also exist contributions from other Higgs productions, VBF and VH with yield rates (ggH:VBF:VH $\sim$ 70:20:10$\%$). However as the leading jets from VBH are most likely quark jets opposed to the gluonic leading jets in the ggH+jets, we checked that the VBF can be easily separated (see Appendix.~\ref{app:vbf}) from the ggH+jets by tagging gluonic central leading jet in addition to the forward jet tagging for VBF. As for the VH which also has quark jet like leading ISRs according to the same argument with the V+jets, it can have additional selection criteria \citep{Ellis:2009me,Thaler:2010tr} for identifying jets from hadronically decaying vector bosons. 
In this regard, to demonstrate the main idea without making event selection scheme too complicated, we simply consider the ggH+jets as the only signal versus the V+jets as the main background in this analysis, without loss of consistency in applying the flavor information for discrimination of gluon-jet rich ggH signal from general quark-jet rich backgrounds.

%
\section{Multivariate Analysis and Result}\label{sec:mva_result}

We use a set of jet substructure variables \cite{Larkoski:2017jix}, say $S^{\, \text{jet}}$, in our analysis as the following,
\begin{itemize}
\item $S^{\,\text{jet}}$ $\equiv$ $\{$ $n_{\text{tk}}$ (track multiplicity) \cite{Gallicchio:2011xq}, Girth \cite{Gallicchio:2010dq,Gallicchio:2011xq}, Broadening \cite{Catani:1992jc}, EEC (energy-energy correlation) \citep{Bhattacherjee:2016bpy} with $\beta=0.2$ \cite{Larkoski:2013eya}, RMS-$p_{T}$ \cite{Gallicchio:2011xq} $\}$,
\end{itemize}
which contain the information on jet flavors. It can also be extended to include more raw data, e.g. jet images \cite{Komiske:2016rsd,Cogan:2014oua} for deep learning. Among the five jet substructure variables used, the Girth as the linear radial moment of a jet reflects a fatness/radius of a jet. As gluon jets tend to have more showers and be fatter by the color factor enhancement, $C_{A}$($g \to gg)/C_{F}$($q \to gq$), such a property can be checked in the Girth distribution of the leading jet from ggH and V+jet processes in Fig.~\ref{fig:templates}(b). 

Jet substructure observables have been used to build a {\it jet tagger}, $\text{P}_{q/g}(S^{\,\text{jet}})$, while the kinematic observables, such as reconstructed four-momenta of jets have been used to build an {\it event classifier}, $\text{P}_{\text{S/B}}(\{p^{\,\text{jet}},...\})$.
However, as can be seen from $d^{2}\sigma/dp^{\,\text{jet}}_{T}dy^{\,\text{jet}}$ in Fig.~\ref{fig:lhc_lumi_rapidity}(c), the flavor of a jet can have a correlation with kinematic information depending on the scattering process. This observation motivates us to build $\text{P}_{\text{S/B}}(\{p^{\,\text{jet}},...\} \cup S^{\,\text{jet}})$, rather than a factorized classifier, $\text{P}_{\text{S/B}}(\{p^{\,\text{jet}},...\})\otimes  \text{P}_{q/g}(S^{\,\text{jet}})$.

\begin{figure}[!t]
\centering
\includegraphics[width=0.48\textwidth]{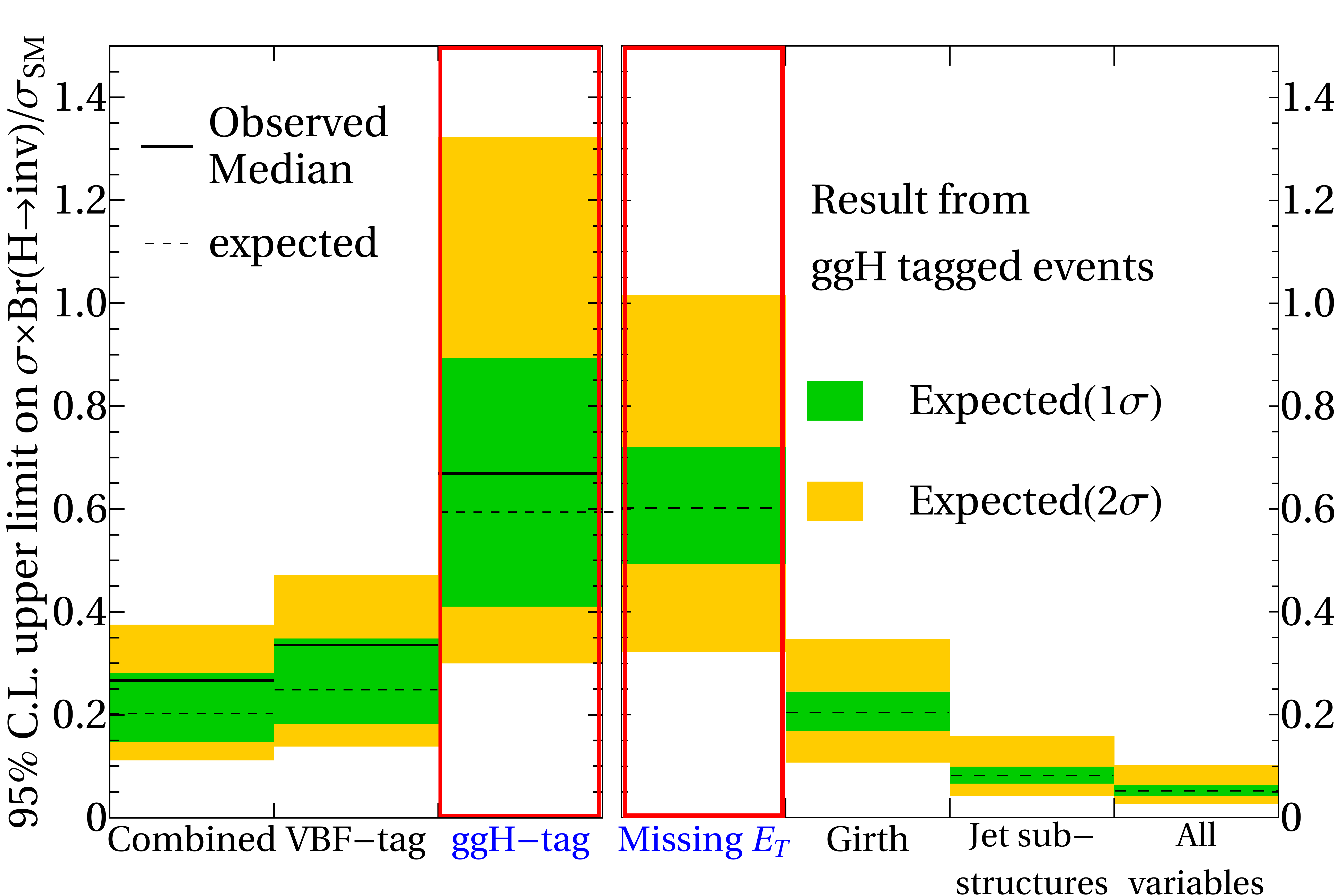}
\caption{Upper limit in 95\% of confidence level(C.L.) on $\sigma \times {\rm Br} (H \to {\rm inv}) \over \sigma_{\rm SM}$ with the integrated luminosity $36\text{fb}^{-1}$. Here we used 5,000 ensembles of pseudo data set which consists of background events only. Left panel shows the results from the experiment \cite{Sirunyan:2018owy}.}
\label{fig:result}
\end{figure}
\begin{table}[!t]
	\centering
	\setlength{\tabcolsep}{0.7mm}
	\renewcommand{\arraystretch}{1.4}
	\scalebox{0.95}{
	\begin{tabular}{  c | c | c | c | c  }
	\hline
	\hline
		$\mathcal{L}$ & $E^{\text{miss}}_{T}$ & Girth & DNN(jet sub) & DNN(all) \\
	\hline
		36$\text{fb}^{-1}$ & $60.2^{+30.0}_{-18.3}\,\%$  & $20.4^{+10.1}_{-5.99}\,\%$ & $8.3^{+4.46}_{-2.55}\,\%$ & $5.2^{+2.83}_{-1.54}\,\%$ \\
	\hline
	\end{tabular}
}
	\caption{Summary on the upper limits in 95\% of confidence level on $\sigma \times \text{BR}(\text{H}\to\text{inv.}) \over \sigma_{\text{SM}}$ for 36 $\fb^{-1}$ at the LHC, for the four template distributions in Fig.~\ref{fig:templates}. The errors were rescaled by the factors projecting the `Missing ET' band to the reference `ggH-tag' band in Fig.~\ref{fig:result}.}
	\label{tab:result}
\end{table}
For S (B) $=$ ggH (V+jets) process, Fig.~\ref{fig:templates} shows the normalized distribution of (a) $E_T^{\text{miss}}$, (b) Girth, (c)  $\text{P}_{\text{S/B}}(S^{\,\text{jet}})$, and (d) $\text{P}_{\text{S/B}}$ $(S^{\,\text{jet}} \cup$  $\{ E_T^{\text{miss}},$ $p_T^{\,\text{jet}},$ $\eta^{\,\text{jet}}\})$. The two event classifiers $\text{P}_{\text{S/B}}$ in Fig.~\ref{fig:templates}(c) and \ref{fig:templates}(d) are obtained by training neural networks with 2-4 layers each with 200-300 nodes with the specified input features. We used one million event samples with Keras \cite{chollet2015keras} for building and training the neural network models. More detailed information about using neural network can be found in Appendix.~\ref{app:dnn}. It is noticeable that the event classifier using the set of jet substructures alone can provide much better separation of signal and background compared to the one $E_T^{\text{miss}}$ as in Fig.~\ref{fig:templates}(c). 
Combining them all we get the best separation as is clearly seen in Fig.~\ref{fig:templates}(d).

The result obtained up to now can be used to discover invisible Higgs decay or put constraints on the invisible branching ratio of Higgs. After selecting the events with the criteria, we performed the profile likelihood ratio test following the procedure in \cite{ATLAS:2011tau} with the four template distributions in Fig.~\ref{fig:templates}.
The likelihood function is given,
\begin{equation}
	\mathcal{L} = \prod_{i=1}^{N_{\text{bin}}}{\frac{\hat{n}^{n_{i}}_{i}  }{n_{i}!} e^{-\hat{n}_{i}}  } \times \frac{1}{2\pi}e^{-\frac{1}{2}\left( \theta^{2}_{s}  + \theta^{2}_{b}\right)},
\end{equation}
where $n_{i}$ is the number of events (or pseudo events) in $i$-th bin, and $\hat{n}_{i}$ is the number of expected events with branching ratio parameter $\mu = \frac{\sigma}{\sigma_{\rm{SM}}}\times BR(h\to \rm{inv})$ (production x-section of invisibly decaying Higgs over the total x-section of the SM Higgs - $\sigma_{\rm{SM}}$), i.e.,
\begin{equation}
	\hat{n}_{i} =\mu  N_{s}P_{s}(i) (1+f_{s})^{\theta_{s}} + N_{b}P_{b}(i) (1+f_{b})^{\theta_{b}}.
\end{equation}
Here the $P_{s (b)}(i)$ is the expected event rate in $i$-th bin, given the total number of events $N_{s(b)}$ survived the cut, and the $\theta_{s(b)}$ in the Poisson and prior probabilities denotes a nuisance parameter associated to the systematic uncertainty $f_{s(b)}$, of signal (background). As a global variation of event rates in signal and background distributions, we tested the $f_{s(b)}$ in 5-20\% \cite{Khachatryan:2016whc}, which changes median of expected upper limit in their 3\% at most, and set $f_{s(b)}$ to 10\%.
Signal cross section (ggH+X) is taken from \cite{deFlorian:2016spz} computed at NNLO+NNLL QCD and NLO EW, and we applied the efficiency on the selection criteria evaluated using simulated event samples, for the fiducial signal yield. For the background process (V+jets) we take both of the cross section and efficiency from our MC simulation of the $Z$+jets, and the fiducial background yield was obtained by a K-factor (1.53) with respect to the $Z$+jet, to take the $W$+jets into account simply, reproducing the expected limit using missing transverse energy \cite{[{(e.g. see Table.4) }]Sirunyan:2017jix}.

For the four profile likelihood ratio tests with/without the new features of jet flavors (`Missing $E_T$', `Girth', `jet substructure variables', and `All variables'), we obtained the upper limits on the branching ratio of invisible Higgs decays in 95\% of confidence level, for the integrated luminosity 36 $\fb^{-1}$ at the LHC, as in Fig.~\ref{fig:result} (right panel), and show them with the existing experimental results (left panel) \cite{Sirunyan:2018owy} - `ggH-tag', `VBF-tag', and `combined', which did not use jet flavor information. As the results in the two red-boxed columns can directly be compared with under the same features and selection criteria, we summarize our results in Table~\ref{tab:result} with the errors rescaled by the correction factors projecting the obtained `Missing $E_T$' band to the reference `ggH-tag' band.
{\it The result shows that the limit on Higgs invisible decays from the ggH can significantly be improved from $60\%$ down to $5\%$ if sub-jet level information of the leading ISR is employed.}
It is interesting that the jet substructures alone provides stronger constraints ($8\%$) than the missing transverse energy in Higgs invisible search. Moreover, combining the features in two kinds, we end up with the best sensitive result ($5\%$) only from the ggH, much lower than the one obtained from VBF ($\sim 20\%$). 

Though more sophisticated understanding and treatment of systematic errors are necessary to obtain a firm number for the expected limit on the Higgs invisible decays, the exercise we did in this paper strongly suggests that 1\% (2\%) precision for Higgs invisible branching ratio at the end of the LHC running with 3 $\ab^{-1}$ (300 $\fb^{-1}$) is a plausible expectation from the gluon fusion solely.
It is also expected that the limit can significantly be improved again if it is combined with the results from VBF and other processes. 

%
\section{Conclusion}\label{sec:conclusion}

We revisited and generalized the property - \textit{the gluon rich leading ISR jets in central rapidity region from gluon fusion Higgs productions versus the quark rich EWVB backgrounds}, and proposed the idea \textit{to improve general Higgs searches produced from ggH by tagging the central gluonic ISR jets}.
	Applying the new method to the searches of invisible decays of Higgs, we showed that the ggH can be the best channel with the improved limit on invisible Higgs decay branching ratio ($60\%\rightarrow 5\%$), significantly exceeding the best limit given by the other channels - VBF ($\sim 20\%$) and VH ($\sim 40\%$). 
The physics and methods in this analysis can also be applied to a broad range of new resonance and Higgs productions induced by gluon fusion, e.g. in search for exotic/rare Higgs decays and di-Higgs productions, concurrently with their irreducible EWVB backgrounds mostly containing quark jet dominant ISRs.

\begin{acknowledgements}
This work was supported by the National Research Foundation of Korea (NRF), Grants No.  2017R1A2B2010749. WC and DL were also supported by the NRF funded by the Ministry of Science and ICT, Grant No. NRF-2017R1C1B2011048. 
\end{acknowledgements}

\clearpage
\begin{appendix}

\begin{widetext}
\section{Loop Functions}\label{app:higgs_loop-functions}
Here we summarised the loop functions relevant for the Higgs production via gluon fusion with a jet at the leading order with finite top-quark mass.
The definition and integral forms are referred to \cite{Ellis:1987xu}.
The two loop functions relevant for $gg \to Hg$, $A_{2}(s,t,u)$  and $A_{4}(s,t,u)$ are defined by,
\begin{eqnarray}
	A_{2}(s,t,u) &=& b_{2}(s,t,u) + b_{2}(s,u,t),  \\
	A_{4}(s,t,u) &=& b_{4}(s,t,u) + b_{4}(t,u,s) + b_{4}(u,s,t). \nn
\end{eqnarray}
Here $b_{2}$ and $b_{4}$ are defined by,
\begin{eqnarray}
	b_{2}(s,t,u) &=& \frac{m^{2}_{t}}{m^{4}_{H}}\left[ \frac{s(u-s)}{u+s} + \frac{2ut(u+2s)}{(u+s)^{2}}\left(W_{1}(t) - W_{1}(m^{2}_{H}) \right) \right.  \\
				&\,& +\left( m^{2}_{t} - \frac{1}{4}s\right)\left( \frac{1}{2}W_{2}(s)+\frac{1}{2}W_{2}(m^{2}_{H}) - W_{2}(t) + W_{3}(s,t,u,m^{2}_{H}) \right) \nonumber \\
				&\,& +s^{2}\left( \frac{2 m^{2}_{t}}{(s+u)^{2}} - \frac{1}{2(s+u)} \right)\left( W_{2}(t) - W_{2}(m^{2}_{H}) \right) \nonumber \\
				&\,& +\left.\frac{ut}{2s}\left( W_{2}(m^{2}_{H})-2W_{2}(t) \right) + \frac{1}{8}\left( s - 12 m^{2}_{t} -\frac{4ut}{s}\right)W_{3}(t,s,u,m^{2}_{H}) \right] \nonumber  \\
	b_{4}(s,t,u) &=& \frac{m^{2}_{t}}{m^{2}_{H}}\left[ -\frac{2}{3} + \left( \frac{m^{2}_{t}}{m^{2}_{H}} - \frac{1}{4} \right)\left( W_{2}(t) - W_{2}(m^{2}_{H}) + W_{3}(s,t,u,m^{2}_{H}) \right)\right], \nonumber
\end{eqnarray}
with $m_{t}$, the mass of top-quark, and $m_{H}$, the mass of Higgs.
The other five light quarks are considered to be massless.
Again, $W_{1}$, $W_{2}$ and $W_{3}$ can be defined as integral forms,
\begin{eqnarray}
	W_{1}(s) &=& 2+ \int_{0}^{1}{dx \ln{\left( 1-x(1-x) \frac{s}{m^{2}_{t}} - i \varepsilon\right)} } \\
	W_{2}(s) &=& 2\int_{0}^{1}{\frac{dx}{x}\ln{\left( 1-x(1-x) \frac{s}{m^{2}_{t}} - i \varepsilon\right)} } \nonumber \\
	W_{3}(s,t,u,v) &=& I_{3}(s,t,u,v) - I_{3}(s,t,u,s) - I_{3}(s,t,u,u) \nonumber \\
	I_{3}(s,t,u,v) &=& \int_{0}^{1}{dx\left( \frac{m^{2}_{t} t}{us}+x(1-x) \right)^{-1}\ln{\left( 1 - x(1-x)\frac{v}{m^{2}_{t}}- i \varepsilon\right)} }. \nonumber
\end{eqnarray}
The other form-factor $A_{5}$ are as follow.
\begin{eqnarray}
	A_{5}(s,t,u) &=& \frac{m^{2}_{t}}{m^{2}_{H}}\left[ 4+ \frac{4s}{t+u}\left( W_{1}(s) - W_{1}(m^{2}_{H})\right) \right. \\
	&\;&\left. \quad  \qquad + \left( 1- \frac{4m^{2}_{t}}{t+u} \right)\left( W_{2}(s)-W_{2}(m^{2}_{H})\right)\right]\nn
\end{eqnarray}
Note that, in the main text, $A_{5}$ has $(\hat{t},\hat{s},\hat{u})$ as its argument, rather than $(\hat{s},\hat{t},\hat{u})$.
This is because the loop function $A_{5}$ was evaluated for $q\bar{q} \to Hg$ rather than $gq \to Hq$.
Under the crossing symmetry, the differential cross section $d\hat{\sigma}_{q\bar{q} \to Hg}/d\hat{t}$ of $q\bar{q} \to Hg $ satisfies
\begin{equation}
	\frac{d\hat{\sigma}_{Hq}}{d\hat{t}}(\hat{s},\hat{t},\hat{u}) = - \frac{N_{c}}{N^{2}_{c}-1}\frac{d\hat{\sigma}_{q\bar{q}\to Hg}}{d\hat{t}}(\hat{t},\hat{s},\hat{u}),
\end{equation}
and we used the same loop function with different argument for $gq \to Hq$ process.

\end{widetext}

\section{Details of Deep Neural Network Structure}\label{app:dnn}

The event classifiers which distinguishes ggH signal from V+jets background is denoted as `$\bf{P}_{\rm{S/B}}(S^{\;\rm{jet}})$' and `$\bf{P}_{\rm{S/B}}(S^{\;\rm{jet}}\cup\{\rm{kin.}\})$'.
$\rm{P}_{\rm{S/B}}(S^{\;\rm{jet}})$ uses 5 jet substructure variables, track multiplicity($n_{\rm{tk}}$, \cite{Gallicchio:2011xq}), girth($G$, \cite{Gallicchio:2011xq,Gallicchio:2010dq}), broadening($B$, \cite{Catani:1992jc}), EEC($C^{\beta}_{1}$, \cite{Bhattacherjee:2016bpy}) with $\beta=0.2$ \cite{Larkoski:2013eya} and RMS-$p_{T}$ \cite{Gallicchio:2011xq}of jet.
$\rm{P}_{\rm{S/B}}(S^{\;\rm{jet}}\cup\{\rm{kin.}\})$ uses $E^{\rm{miss}}_{T}$, $p^{\;\rm{jet}}_{T}$ and $\eta^{\;\rm{jet}}_{T}$ in addition to jet substructure variables used in $\rm{P}_{\rm{S/B}}(S^{\;\rm{jet}})$, so total 8 variables are used.
The definition of jet substructure variables are as the following:
\begin{eqnarray}
    \text{girth : } G &=& \frac{1}{p^{\;\text{jet}}_{T}}\sum_{i \in \{\text{const.} \}}{p^{i}_{T} \left| \Delta \vec{r}_{i} \right|}, \\
    \text{broadening : } B &=& \frac{1}{\sum_{i}{\left| \vec{p}^{\;i}\right| }}\sum_{i}{\left| \vec{p}^{\;i} \times \hat{p}^{\;\text{jet}} \right|}, \\
    &=&  \frac{1}{\sum_{i}{\left| \vec{p}^{\;i}\right| }}\sum_{i}{\left|\vec{k}^{i}_{T} \right|}, \nn \\
    \text{EEC : } C^{\beta}_{1} &=& \frac{1}{\left( \sum_{i}{p^{i}_{T}}\right)^{2} }\sum_{i<j}{p^{i}_{T}p^{j}_{T}\left(\Delta R_{ij}\right)^{\beta}}, \\
    \text{RMS-}p_{T}\text{ : }  \sqrt{\langle p^{2}_{T} \rangle} &=& \frac{1}{p^{\;\text{jet}}_{T}}\sqrt{\frac{1}{n_{\text{tk}}}\sum_{i}{\left(p^{i}_{T}\right)^{2}}}
\end{eqnarray}
where $\{\rm{const.}\}$ means the set of constituents of a jet.

\begin{figure*}[!htbp]
	\centering
	\includegraphics[width=0.3\textwidth]{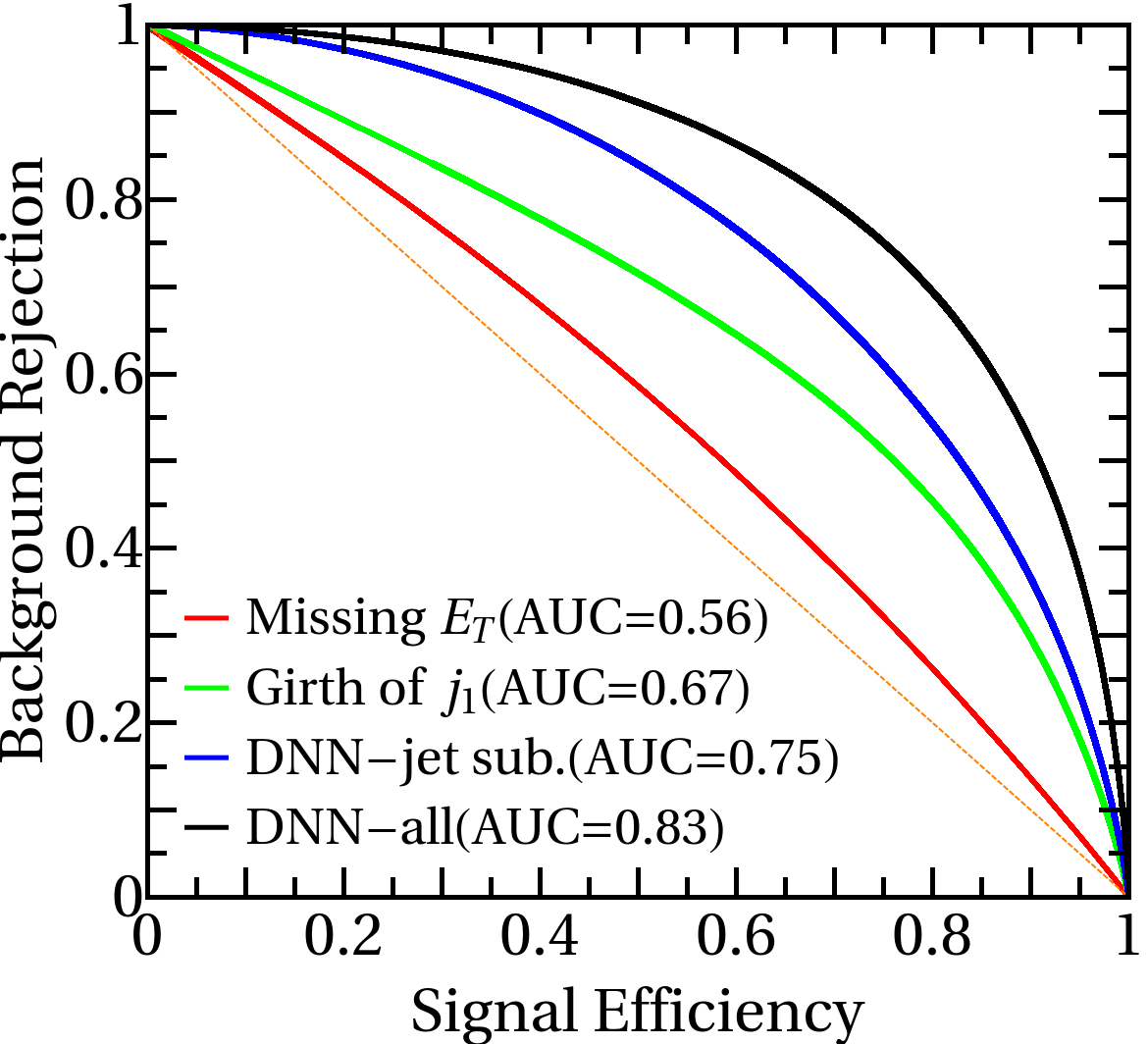} \hspace{2cm}
	\includegraphics[width=0.3\textwidth]{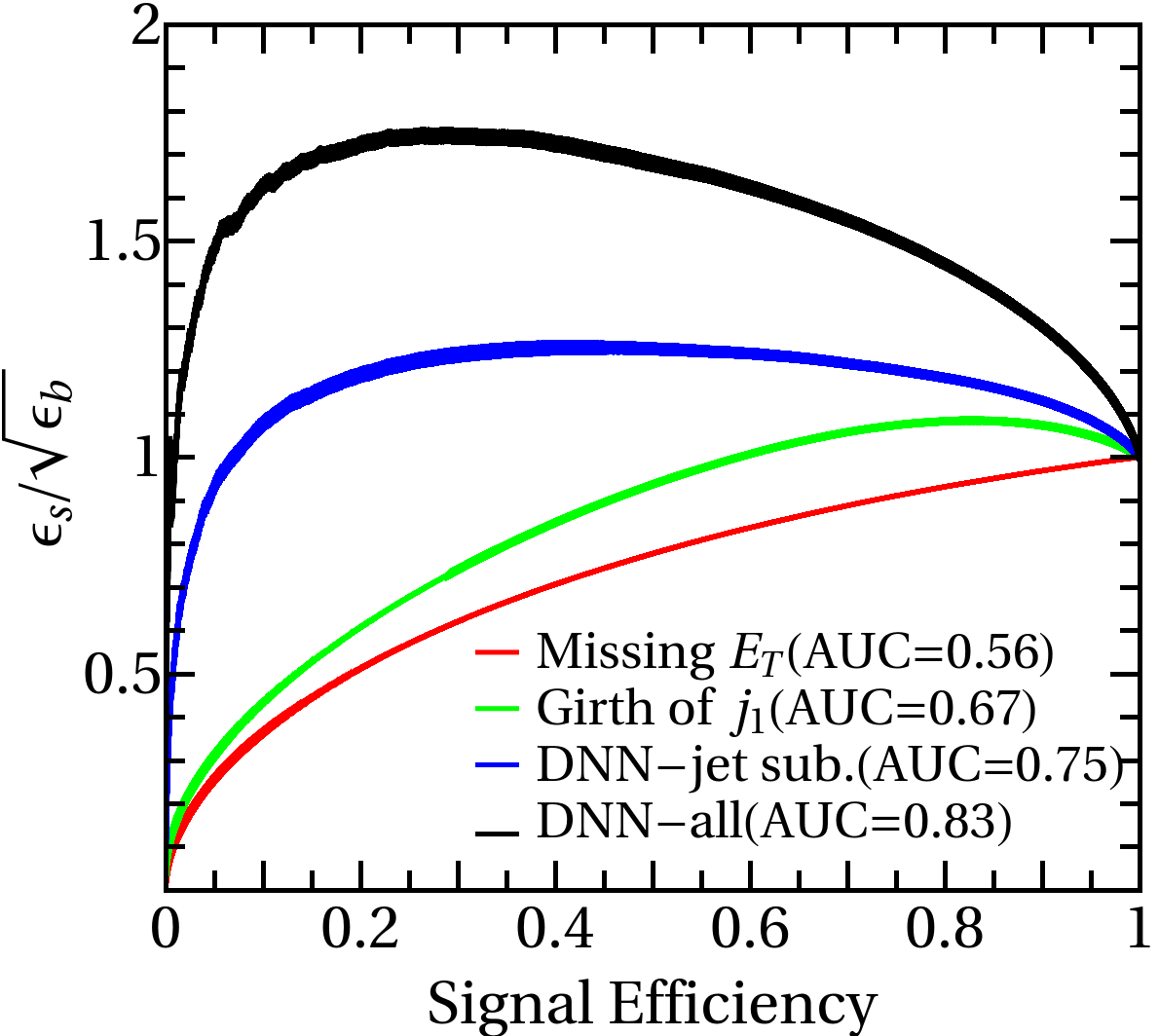}
	\caption{Receiver operating characteristic (ROC) curves (left) and $\epsilon_{\text{sig}}/\sqrt{\epsilon_{\text{bg}}}$ curves (right) derived from respective ROC curves. In each panel, ROC derived from $E^{\rm{miss}}_{T}$ distribution in drawn with red, and one from girth is green. $\rm{P}_{\rm{S/B}}(S^{\;\rm{jet}})$ and $\rm{P}_{\rm{S/B}}(S^{\;\rm{jet}}\cup\{\rm{kin.}\})$ are drawn in blue and black, respectively.} 
	\label{fig:inv_higgs_roc}
\end{figure*}

The specification of the neural network classifiers - $\rm{P}_{\rm{S/B}}(S^{\;\rm{jet}})$ and $\rm{P}_{\rm{S/B}}(S^{\;\rm{jet}}\cup\{\rm{kin.}\})$, is summarized in Table~\ref{tab:dnn_spec} especially on its structure, hyper-parameters and training prescriptions.
In addition to the distribution of neural network outputs in the main text, here the receiver operating characteristic (ROC) curves and related $\epsilon_{\text{sig}}/\sqrt{\epsilon_{\text{bg}}}$ are shown in Fig.~\ref{fig:inv_higgs_roc},

\begin{table*}[thb]
	\centering
	{\renewcommand{\arraystretch}{1.2}
	\begin{tabular}{| l | >{\centering\arraybackslash}m{12em} | >{\centering\arraybackslash}m{10em} |}
	\toprule
							&$\bf{P}_{\rm{S/B}}(S^{\;\rm{jet}})$ & $\bf{P}_{\rm{S/B}}(S^{\;\rm{jet}}\cup\{\rm{kin.}\})$ \\
	\hline	Training data & \multicolumn{2}{c |}{1M(0.5M each for sig/bg)} \\
	\hline	Validation data & \multicolumn{2}{c |}{1M(0.5M each for sig/bg)} \\
	\hline	Preprocessing & \multicolumn{2}{c |}{Standard-scaler} \\
	\hline	NN package & \multicolumn{2}{c |}{Keras \cite{chollet2015keras} with TensorFlow backend} \\
	\hline	NN structure & \multicolumn{2}{c |}{ Fully-connected feed-forwarding (FF) layers} \\
	\hline	Normalisation & \multicolumn{2}{c |}{Batch normalization \cite{ioffe2015batch} } \\
	\hline	Drop out & 30\% & 10\% \\
	\hline	NN structure & {2 hidden FF layers with 300 nodes each} & {4 hidden FF layers with 200 nodes each} \\
	\hline Optimiser & \multicolumn{2}{c |}{Adam \cite{kingma2017adam}} \\
	\hline Loss function & \multicolumn{2}{c |}{Categorical cross-entropy} \\
	\hline Learning rate & 0.001 & 0.001 \\
	\hline Mini-batch size & \multicolumn{2}{c|}{50,000} \\
	\hline Activation function & \multicolumn{2}{c|}{ReLU for intermediate layers, soft-max for output layer} \\
	\hline Initialisation & \multicolumn{2}{c|}{He \cite{he2015delving}} \\
	\hline
	\end{tabular}
	}
	\caption{DNN model specification and training prescriptions used for this study.}
	\label{tab:dnn_spec}
\end{table*}

\section{Separation of Vector-Boson-Fusion Higgs Production}\label{app:vbf}

In this section we shortly discuss about the discrimination of vector-boson-fusion(VBF) Higgs production mechanism from another Higgs production mechanism, ggH, and its background, Drell-Yan(DY) process.
It is well known that the VBF process has characteristic two jets with large angular separation between them($|\Delta \eta_{\text{jj}}|$) and large invariant mass of them($m_{\text{jj}}$).
This handle is used to separate the VBF from the other processes, ggH and DY.
On top of that, there is additional handle, which is the parton contents of jets.
Note that the two jets of VBF processes are mostly quark jets, while the leading jets of ggH in central region are mainly gluonic.
Focusing on the leading jet flavor, therefore, does separate the VBF process from ggH process.
At the same time, the sub-leading jets from DY process are more likely to be gluonic, due to parton luminosity.
Hence VBF process can be also separated from DY process by observing the parton contents of the sub-leading jets.

\begin{figure*}[!htbp]
	\centering
	\includegraphics[width=0.9\textwidth]{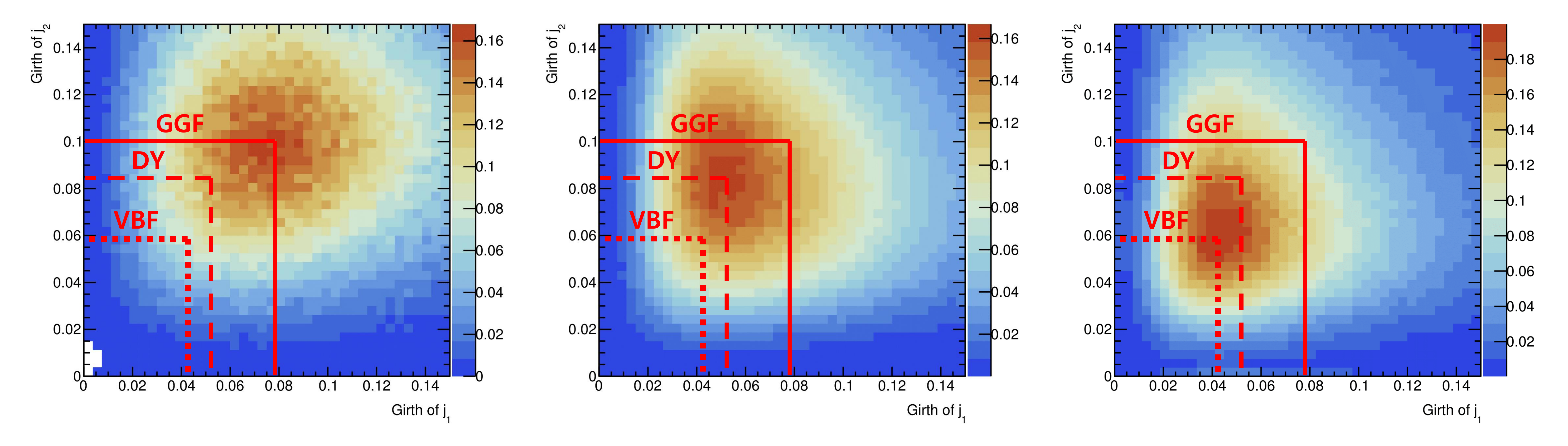}
	\caption{The scatter plots of girth of leading jets ($x$-axis) versus girth of sub-leading jets ($y$-axis). Three plots came from different processes, ggH (left), DY (middle) and VBF (right). Red lines in plots indicate the peak position from three distributions. The solid red line points the peak from ggH process, and dashed line for DY  and dotted line for VBF. Details of event samples for this plot can be found in the text.}
	\label{fig:girth_2d}
\end{figure*}

The difference in parton contents of associated jets can be seen in the two-dimensional distribution of leading jet girth, and sub-leading jet girth as in Fig.~\ref{fig:girth_2d}.
The event samples generated at 14 TeV with aMC@NLO \cite{Alwall:2014hca} at NLO in QCD with $m_{t}\to\infty$ limit are used for this plot.
CT10NLO \cite{Lai:2010vv} PDF and Higgs Characterisation \cite{Artoisenet:2013puc} model are used for the simulation.
Events are showered using Pythia8 \cite{Sjostrand:2014zea} and 
merged via FxFx \cite{Frederix:2012ps} scheme with $Q_{\text{cut}} = 40 \text{ GeV}$. 
The jet clustering is done with anti-$k_{T}$ algorithm with $\Delta R = 0.4$ 
using FastJet \cite{Cacciari:2011ma}, for $p_{T}^{\text{jet}} \ge 30\text{ GeV}$.
The fast detector simulation is done with Delphes \cite{deFavereau:2013fsa}.

In Fig.~\ref{fig:girth_2d}, the peaks of 2-dimensional girth distributions from each process placed away from the peaks of the others due to different quark/gluon composition of corresponding jets as we expected. 
\end{appendix}
\clearpage
\bibliography{reference}
\end{document}